\documentclass[10pt,twocolumn]{article}
\usepackage{latex8}
\usepackage{color,times}
\usepackage{amsfonts,epsfig,amsmath,psfig}

\begin{document}

\newcommand{\cris}

\newcommand{\C}{{\cal C}}

\newcommand{\dalpha}{{\rm d}\alpha}

\def\reals{\mathbb{R}}
\def\naturals{\mathbb{N}}

\def\eg{e.g.\ }
\def\ie{i.e.,\ }

\def\as{a.s.\ }
\def\whp{w.h.p.}
\def\Whp{W.h.p.}

\newcommand{\fo}[2]{\ensuremath{{F}_{#1}(n,#2)}}
\def\var{{\rm var}}
\def\ex{{\bf E}}

\newtheorem{definition}{Definition}
\newtheorem{lemma}{Lemma}
\newtheorem{claim}{Claim}
\newtheorem{theorem}{Theorem}
\newtheorem{corollary}{Corollary}
\newtheorem{conjecture}{Conjecture}

\newcommand{\ra}{\rightarrow}

\newcommand{\dima}{}
\newcommand{\poor}{}

\newcommand{\e}{{\rm e}}
\newcommand{\eps}{\epsilon}
\newcommand{\beq}{\begin{equation}}
\newcommand{\eeq}{\end{equation}}
\newcommand{\bea}{\begin{eqnarray*}}
\newcommand{\eea}{\end{eqnarray*}}
\newcommand{\sub}{\scriptsize}
\newcommand{\mat}{\left(\!\!\begin{array}{cc}}
\newcommand{\rix}{\end{array}\!\!\right)}
\newcommand{\ord}{{O}}
\newcommand{\bra}{\langle}
\newcommand{\ket}{\rangle}
\newcommand{\p}{\partial}
\newcommand{\Z}{{\Bbb Z}}
\newcommand{\dm}{{\rm d}m}
\newcommand{\ds}{{\rm d}s}
\newcommand{\dt}{{\rm d}t}
\newcommand{\dv}{{\rm d}v}
\newcommand{\dx}{{\rm d}x}
\newcommand{\dy}{{\rm d}y}
\newcommand{\dw}{{\rm d}w}
\newcommand{\db}{{\rm d}b}
\newcommand{\dc}{{\rm d}c}
\newcommand{\du}{{\rm d}u}
\newcommand{\dbeta}{{\rm d}\beta}
\newcommand{\dgamma}{{\rm d}\gamma}
\newcommand{\ddelta}{{\rm d}\delta}
\def\limninf{\lim_{n \rightarrow \infty}}

\renewcommand{\d}{{\delta}}
\newcommand{\amax}{\alpha_{\max}}
\newcommand{\bmax}{\beta_{\max}}
\newcommand{\gamax}{\gamma_{\max}}
\newcommand{\gmax}{g_{\max}}
\newcommand{\za}{\vec{\zeta}}
\newcommand{\zamax}{\vec{\zeta}_{\max}}
\newcommand{\HTC}{{\rm H}}
\newcommand{\NAESAT}{{\rm N}}

\title{The Asymptotic Order of the Random $k$-SAT Threshold}

\author{
Dimitris Achlioptas \\
Microsoft Research, Redmond, Washington \\
{\tt optas@microsoft.com} \and
Cristopher Moore\thanks{Supported by NSF grant PHY-0071139, the Sandia
University Research Program, and Los Alamos National Laboratory.} \\
Computer Science Department, University of New Mexico, Albuquerque \\
and the Santa Fe Institute, Santa Fe, New Mexico \\
{\tt moore@cs.unm.edu}
}

\maketitle
\thispagestyle{empty}

\begin{abstract}
  Form a random $k$-SAT formula on $n$ variables by selecting
  uniformly and independently $m=rn$ clauses out of all $2^k
  \binom{n}{k}$ possible $k$-clauses. The {\em Satisfiability
    Threshold Conjecture\/} asserts that for each $k$ there exists a
  constant $r_k$ such that, as $n$ tends to infinity, the probability
  that the formula is satisfiable tends to 1 if $r<r_k$ and to 0 if
  $r>r_k$. It has long been known that $2^k / k < r_k < 2^k$. We prove
  that $r_k > 2^{k-1} \ln 2 - d_k$, where $d_k \rightarrow (1+\ln 2)/2$.
  Our proof also allows a blurry glimpse
  of the ``geometry'' of the set of satisfying truth assignments.
\end{abstract}

\Section{Introduction}

Satisfiability has received a great deal of study as the canonical
NP-complete problem.  In the last twenty years some of this work
has been devoted to the study of randomly generated formulas and
the performance of satisfiability algorithms on them.  Among the
many proposed distributions for generating satisfiability
instances, random $k$-SAT  has received the lion's share of
attention.

For some canonical set $V$ of $n$ Boolean variables, let
$C_k=C_k(V)$ denote the set of all $2^k \binom{n}{k}$ possible
disjunctions of $k$ distinct, non-complementary literals from $V$
\mbox{($k$-clauses)}. A random $k$-SAT formula $F_k(n,m)$ is
formed by selecting uniformly, independently, and with replacement
$m$ clauses from $C_k$ and taking their
conjunction\footnotemark[1].  We will be interested in random
formulas as $n$ grows. In particular, we will say that a sequence
of random events ${\mathcal E}_n$ occurs with high probability
(w.h.p.) if $\lim_{n \ra \infty} \Pr[{\mathcal E}_n]=1$.

There are at least two reasons for the popularity of random
$k$-SAT. The first reason is that while random $k$-SAT instances
are trivial to generate they appear very hard to solve, at least
for some values of the distribution parameters.  The second reason
is that the underlying formulas appear to enjoy a number of
intriguing mathematical properties, including 0-1 laws and a form
of expansion.

\footnotetext[1]{In fact, our discussion and results hold in all
common models for random $k$-SAT, \eg when clause replacement is
not allowed and/or when each $k$-clause is formed by selecting $k$
literals uniformly at random with replacement.}

The mathematical investigation of random $k$-SAT began with the
work of Franco and Paull~\cite{FrPa}. Among other results, they
observed that $F_k(n,m=rn)$ is \whp\ unsatisfiable if $r > 2^k \ln
2$.  To see this, fix any truth assignment and observe that a
random $k$-clause is satisfied by it with probability $1-2^{-k}$.
Therefore, the expected number of satisfying truth assignments of
$F_k(n,m=rn)$ is $[2(1-2^{-k})^r]^n=o(1)$ for $r > 2^k \ln 2$.
Shortly afterwards, Chao and Franco~\cite{ChFrUC} complemented
this result by proving that for all $k \geq 3$, if $r < 2^k/k$
then the following linear-time algorithm, called {\sc Unit Clause
(uc)}, finds a satisfying truth assignment with probability at
least $\varepsilon=\varepsilon(r)>0$:
\begin{center}
If there exist unit clauses, pick one randomly and satisfy it;
else pick a random unset variable and set it to 0.
\end{center}

A seminal result in the area was established a few years later by
Chv\'{a}tal and Szemer\'{e}di~\cite{ChvSz}. Extending the work of
Haken~\cite{Haken} and Urquhardt~\cite{urq} they proved the
following: for all $k \geq 3$, if $r > 2^k \ln 2$, then  \whp\
$\fo{k}{rn}$ is unsatisfiable   and every resolution proof of its
unsatisfiability must contain at least $(1+\eps)^n$ clauses, for
some $\eps = \eps(k,r)>0$.

Random $k$-SAT owes a lot of its popularity to the experimental
work of Selman, Mitchell and Levesque~\cite{MSL} who considered
the performance of a number of practical algorithms on random
$3$-SAT instances. Across different algorithms, their experiments
consistently drew the following picture: for $r <4$, a satisfying
truth assignment can be found easily for almost all formulas; for
$r>4.5$, almost all formulas are unsatisfiable; and for $r \approx
4.2$, a satisfying truth assignment can be found for roughly half
the formulas, while the observed computational effort is
maximized. The following conjecture, formulated independently by a
number of researchers, captures the suggested 0-1 law:
\medskip

\noindent {\bf Satisfiability Threshold Conjecture} {\em For each
$k \geq 2$, there exists a constant $r_k$ such that}
\[
\limninf \Pr[F_k(n,rn)\mbox{ is satisfiable}]=
\begin{cases}
1 & \mbox{if $r<r_k$}\\
0 & \mbox{if $r>r_k$} \enspace .
\end{cases}
\]

The conjecture was settled early on for the linear-time solvable
case $k=2$: independently, Chv\'{a}tal and Reed~\cite{mick},
Fernandez de la Vega~\cite{vega2sat}, and Goerdt~\cite{GoTU}
proved $r_2=1$. For $k \geq 3$, neither the value nor the
existence of $r_k$ have been established. Friedgut~\cite{frie},
though, has proved the existence of a critical {\em sequence\/}
$r_k(n)$ around which the probability of satisfiability goes from
1 to 0. In the following, we will take the liberty of writing $r_k
\geq r^*$ if \fo{k}{rn} is satisfiable \whp\ for all $r < r^*$
(and analogously for $r_k \leq r^*$).

Chv\'{a}tal and Reed~\cite{mick}, besides proving $r_2=1$, gave
the first lower bound for $r_k$, strengthening the
positive-probability result of~\cite{ChFrUC}. In particular, they
considered a generalization of {\sc uc}, called {\sc sc}, which in
the absence of unit clauses satisfies a random literal in a random
2-clause (and in the absence of 2-clauses satisfies a uniformly
random literal). They proved that for all $k\geq 3$, if $r < (3/8)
2^k/k$  then {\sc sc} finds a satisfying truth assignment \whp

In the last ten years, the satisfiability threshold conjecture has
received attention in theoretical computer science, mathematics
and, more recently, statistical physics.  A large fraction of this
attention has been devoted to the first computationally
non-trivial case, $k=3$ and a long series of
results~\cite{BFU,ChFrGUC,FrSu,pi,326,KKL_talk,dub_announ,Kap,JSV,KKKS,KMPS,EMFV,FrPa}
has narrowed the potential range of $r_3$. Currently this is
pinned between $3.42$ by Kaporis, Kirousis and
Lalas~\cite{KKL_talk} and $4.506$ by Dubois and
Boufkhad~\cite{dub_announ}. All upper bounds for $r_3$ come from
probabilistic counting arguments, refining the idea of counting
the expected number of satisfying truth assignments. All lower
bounds on the other hand have been algorithmic, the refinement
lying in considering progressively more sophisticated algorithms.

Unfortunately, for general $k$, neither of these two approaches
above has helped narrow the asymptotic gap between the upper and
lower bounds for $r_k$. The known techniques improve upon $r_k
\leq 2^k \ln 2$ by a small additive constant, while the best lower
bound, comes from Frieze and Suen's~\cite{FrSu} analysis of a full
generalization of~{\sc uc}:
\begin{center} Satisfy a random literal in a random shortest clause.
\end{center}
This gives $r_k \geq c_k 2^k/k$ where $\lim_{k \ra \infty} c_k =
1.817\ldots$

If one chooses to live unencumbered by the burden of mathematical
proof, then a powerful non-rigorous technique of statistical
physics known as the ``replica trick'' is available.  So far,
predictions based on the replica trick have exhibited a strong
(but not perfect) correlation with the (empirically observed)
truth.  Using this technique, Monasson and Zecchina~\cite{MZRK}
predicted $r_k \simeq 2^k \ln 2$. Like most arguments based on the
replica trick, their argument is mathematically sophisticated but
far from being rigorous.

If one indeed believes that the correct answer lies closer to the
upper bound (for whatever reason) then analyzing more
sophisticated satisfiability algorithms is an available option.
Unfortunately, after a few steps down this path one is usually
forced to choose between rather naive algorithms, which can be
analyzed, or more sophisticated algorithms that might get closer
to the threshold, but are much harder to analyze. In particular,
the lack of progress over $c\,2^k/k \leq r_k \leq 2^k \ln 2$ in
the last ten years suggests the possibility that no (naive)
algorithm can significantly improve the lower bound. At the same
time, it is clear that proving lower bounds by analyzing
algorithms is doing ``more than we need'': we not only get a proof
that a satisfying assignment exists but an explicit procedure for
finding one.

In this paper, we eliminate the asymptotic gap for $r_k$ by using
the ``second moment'' method. Employing such a non-constructive
argument allows us to overcome the limitations of current
algorithmic techniques or, at least, of our capacity to analyze
them. At the same time, not pursuing some particular satisfying
truth assignment affords us a first, blurry glimpse of the
``geometry'' of the set of satisfying truth assignments. Our main
result is the following.
\begin{theorem}\label{thm:ksat}
For all $k \geq 2$, $r_k > 2^{k-1} \ln 2 - d_k$, where $d_k
\rightarrow (1+\ln 2)/2$.
\end{theorem}

As we will see shortly, a straightforward application of the
second moment method to random $k$-SAT fails rather dramatically:
if $X$ denotes the number of satisfying truth assignments, then
$\ex[X^2]>(1+\epsilon)^n \,\ex[X]^2$ for any $r>0$. To prove
Theorem~\ref{thm:ksat} it will be crucial to focus on those
satisfying truth assignments {\em whose complement is also
satisfying}. Observe that this is equivalent to interpreting
$F_k(n,m)$ as an instance of Not All Equal (NAE) $k$-SAT, where a
truth assignment is NAE-satisfying if every clause contains at
least one satisfied literal {\em and\/} at least one unsatisfied
literal.

Analogously to random $k$-SAT, it is trivial to show that if $r>
2^{k-1} \ln 2 - (\ln 2)/2$ then \whp\ $F_k(n,m=rn)$ has no
NAE-satisfying truth assignments since their expected number is
$o(1)$. We match this within an additive constant.
\begin{theorem}
\label{thm:naeksat} There exists a sequence $t_k \rightarrow 1/2$
such that if
\[
r < 2^{k-1} \ln 2 - (\ln 2)/2 - t_k
\]
then w.h.p.\ $F_k(n,rn)$ is NAE-satisfiable.
\end{theorem}

Theorem~\ref{thm:ksat} follows trivially from
Theorem~\ref{thm:naeksat} since any NAE-satisfying assignment is
also a satisfying assignment. Our method actually yields an
explicit lower bound for the random NAE $k$-SAT threshold for each
value of $k$ as the solution to a transcendental equation (yet one
without an attractive closed form, hence
Theorem~\ref{thm:naeksat}). It is, perhaps, worth comparing our
lower bound for the NAE $k$-SAT threshold with the upper bound
derived using the technique of~\cite{KKKS} for small values of
$k$. Even for $k=3$, our lower bound is competitive with the best
known lower bound of $1.514$, obtained by analyzing a
generalization of {\sc uc} that minimizes the number of unit
clauses~\cite{mysoda}. For larger $k$, the gap between the upper
\mbox{and the lower bound rapidly converges to $\approx 1/4$.}

\begin{table}[h]\label{tab:val}
\begin{center}
$ \begin{array}{c|ccccc}
 k              &  3    & 5         &  7        &  10       &    12\\   \hline
\mbox{Lower}    & 3/2   & 9.973     & 43.432    & 354.027 & 1418.712 \\
\mbox{Upper}    & 2.214 & 10.505    & 43.768    & 354.295 & 1418.969 \\
\end{array}
$
\end{center}
\begin{center}
Table 1. Bounds for the random NAE $k$-SAT threshold.
\end{center}
\end{table}
Recently, and independently of our work, Frieze and
Wormald~\cite{friezewormald} showed that another way to
successfully apply the second moment to random $k$-SAT is to let
$k$ grow with $n$. In particular, let \mbox{$\omega=k-\log_2 n
\rightarrow \infty$}, let $m_0=-\frac{n \ln
2}{\ln(1-2^{-k})}=(2^k+O(1))n \ln 2$ and let
$\epsilon=\epsilon(n)>0$ be such that $\epsilon n \ra \infty$.
Then, $F_k(n,m)$ is \whp\ satisfiable if $m < (1-\epsilon)m_0$ but
\whp\ unsatisfiable if $m
> (1+\epsilon)m_0$.

\mbox{}

We prove Theorem~\ref{thm:naeksat} by applying the following
version of the second moment method (see Exercise 3.6
in~\cite{ramr}).
\begin{lemma}\label{lemma:sec}
For any non-negative random variable $X$,
\begin{equation}\label{eq:second}
 \Pr[X > 0] \,\ge\, \frac{\ex[X]^2}{\ex[X^2]} \enspace .
\end{equation}
\end{lemma}
In particular, let $X\geq 0$ be the number of NAE-satisfying
assignments of $F_k(n,m=rn)$.  We will prove that for all $\eps >
0$ and all $k \geq k_0(\eps)$, if $r \leq 2^{k-1} \ln 2 -
\frac{\ln 2}{2} - \frac{1}{2} - \eps$ then there exists some
constant $C=C(k)$ such that
\[
\ex[X^2] < C \times \ex[X]^2 \enspace .
\]
By Lemma~\ref{lemma:sec}, this implies
\[
\Pr[X>0] \geq \Pr[F_k(n,rn) \mbox{ is NAE-satisfiable}] > 1/C
\enspace .
\]
To get Theorems~\ref{thm:ksat} and \ref{thm:naeksat} we boost this
positive probability to $1-o(1)$ by employing the following
corollary of the aforementioned non-uniform threshold for random
$k$-SAT~\cite{frie} (and its analogue for random NAE $k$-SAT):
\begin{corollary}\label{cor:boost}
  If $\liminf_{n \ra \infty} \Pr[F_k(n,r^*n) \mbox{ is satisfiable}] >
  0$, then $F_k(n,rn)$ is satisfiable \whp\ for $r < r^*$.
\end{corollary}

In the next section we give some intuition on why the second
moment method fails when $X$ is the number of satisfying truth
assignments, and how letting $X$ be the number of NAE-satisfying
assignments rectifies the problem.  In Section~\ref{sec:new} we
give some related general observations and point out potential
connections to statistical physics. We lay the groundwork for
bounding $\ex[X^2]$ in Section~\ref{ground}. The actual bounding
happens in Section~\ref{sec:gleas}.  We conclude with some
discussion in Section~\ref{sec:conc}.

 \Section{The second moment method}\label{sec:crapola}

\SubSection{Random $k$-SAT}

Let $X$ denote the number of satisfying assignments of $F_k(n,m)$.
Since $X$ is the sum of $2^n$ indicator random variables,
linearity of expectation implies that to bound $\ex[X^2]$ we can
consider all $4^n$ ordered pairs of truth assignments and bound
the probability that both assignments in each pair are satisfying.
It is easy to see that, by symmetry, for any pair of truth
assignments $s,t$ this probability depends only on the number of
variables assigned the same value by $s$ and $t$, \ie their
overlap. Thus, we can write $\ex[X^2]$ as a sum with $n+1$ terms,
one for each possible value of the overlap $z$, the $z$th such
term being: $2^n$ (counting over $s$) $\times$ an ``entropic''
$\binom{n}{z}$ factor (counting overlap locations) $\times$ a
``correlation'' factor measuring the probability that truth
assignments $s,t$ having overlap $z$ are both satisfying.

Now, as we saw earlier, $\ex[X] = [2 (1-2^{-k})^r]^n = c^n$. Thus,
if $r$ is such that $c < 1$, then $\Pr[X>0] \leq \ex[X] = o(1)$
and we readily know that $F_k(n,rn)$ is \whp\ unsatisfiable. (Note
that $\Pr[X>0]=o(1)$ even when $c = 1$ since the naive upper bound
is not tight.) Therefore, we are only interested in the case where
$\ex[X^2] \geq \ex[X]^2 = (1+\varepsilon)^n$ for some $\varepsilon
= \varepsilon(r) > 0$. Since the sum defining $\ex[X^2]$ has only
$n+1$ terms we see that, up to polynomial factors, $\ex[X^2]$ is
equal to the contribution of the term maximizing the
``entropy-correlation'' product.

Observe, now, that if $z=n/2$, then the probability that $s$ and
$t$ are both satisfying is the square of the probability that one
of them is. To see this take $s$ to be, say, the all 0s assignment
and consider the set of clauses this precludes from being in the
formula. Thus, for truth assignments that overlap on $n/2$ bits,
the events of being satisfying are independent. Therefore, up to
polynomial factors, $\ex[X]^2$ is equal to the $z=n/2$ term of the
sum defining $\ex[X^2]$.

From the above discussion, letting $\alpha=z/n$, we see  that if
the entropy-correlation factor is maximized at some $\alpha \neq
1/2$ then the second moment method fails. On other other hand, as
we will see, if the maximum does indeed occur at $\alpha=1/2$,
then the polynomial factors cancel out and the ratio
$\ex[X^2]/\ex[X]^2$ is bounded by a constant independent of $n$,
implying that in that case $\Pr[X>0]> 1/C$.

\begin{figure}
  \centerline{\psfig{figure=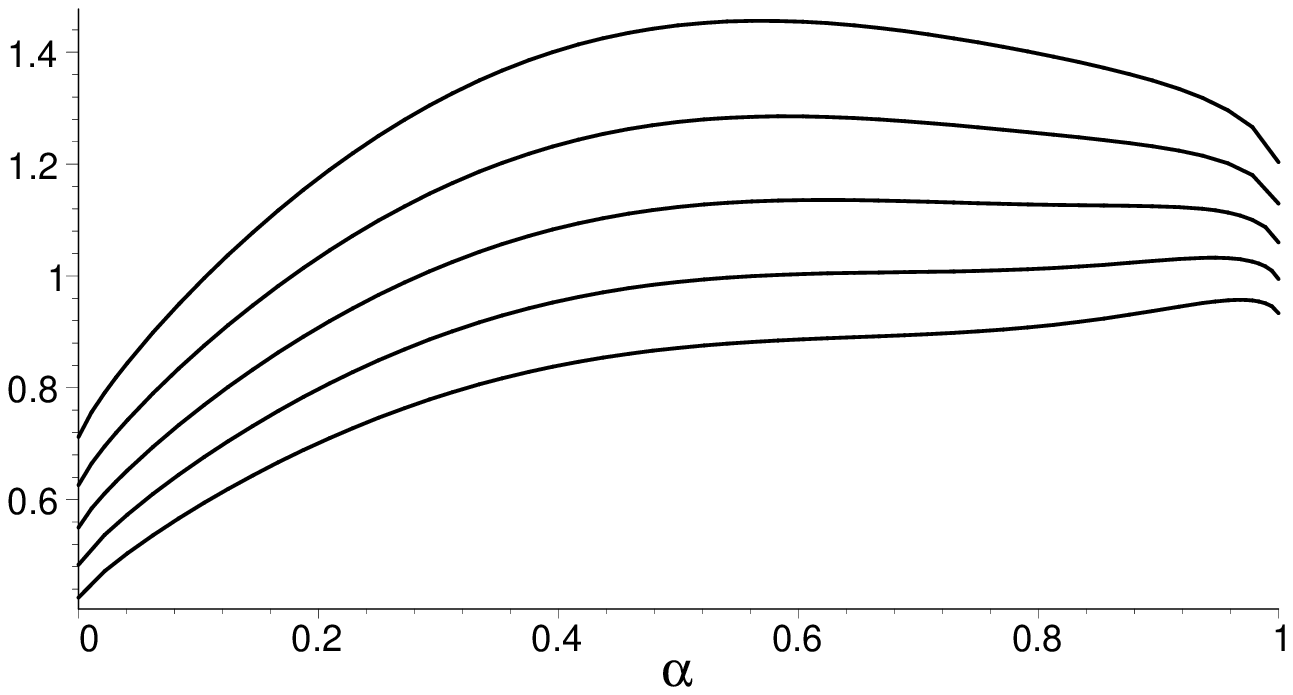,height=1.5in,width=2in}}
  \centerline{$k=5$, $r=16,18,20,22,24$ (top to bottom)}
\bigskip
Figure 1.  {\em The $n$th root of\/ $2 \times$the
entropy-correlation product for $k$-SAT as a function of the
overlap $\alpha=z/n$.}
\end{figure}

\vspace*{-0.6cm}

With these observations in mind, in Fig.~1 we plot the $n$th root
of each of the $n+1$ terms contributing to $\ex[X^2]$ as a
function of $\alpha=z/n$ for $k=5$ and different values of $r$.

Unfortunately, we see that for all values of $r$ considered the
maximum lies to the right of $\alpha=1/2$. The reason for this is
that the correlation factor for $k$-SAT is strictly increasing
with $\alpha=z/n$.  For instance, as we saw above, if $s$ is
satisfying and $t$ has an overlap of $z=n/2$ with $s$, then the
conditional probability that $t$ is also satisfying equals its
{\em a priori} value $(1-1/2^k)^m$. But if $z$ decreases, say, to
$0$ then the conditional probability that $t$ is satisfying
decreases to $(1-1/(2^k-1))^m$, penalizing $t=\overline{s}$
exponentially and making it the least likely assignment to be
satisfying.

This asymmetry in the correlation factor implies that for all
$r>0$ its product with the (symmetric) entropy factor is maximized
at some $\alpha>1/2$. Therefore, $\ex[X^2]$ is greater than
$\ex[X]^2$ by an exponential factor for all $r>0$, and
Lemma~\ref{lemma:sec} fails to give any non-trivial lower bound.
To have any hope of getting a lower bound by the second method we
need to consider a set of satisfying assignments for which the
derivative of the correlation factor at $1/2$ is zero.

\SubSection{Random NAE $k$-SAT}

One attractive feature of the second moment method is that we are free
to apply it to any random variable $X$ such that $X>0$ implies that
$F_k(n,m)$ is satisfiable. In particular, we can refine our earlier
application of the method by focusing on any subset of the set of
satisfying assignments.

Considering only assignments that are NAE-satisfying --- or,
equivalently, whose complement is also satisfying --- makes the
correlation factor symmetric around $\alpha = 1/2$ as  twin
satisfying assignments $s$ and $\overline{s}$ provide an equal
``tug'' to every other truth assignment $t$.  As a result, we
always have a local extremum at $\alpha=1/2$ since both the
correlation factor and the entropy are symmetric around it.
Moreover, since the entropic term is independent for $r$, this
extremum is a local maximum for sufficiently small $r$. Whenever
this is also the {\em global} maximum, the second moment succeeds.

\begin{figure}
  \centerline{\psfig{figure=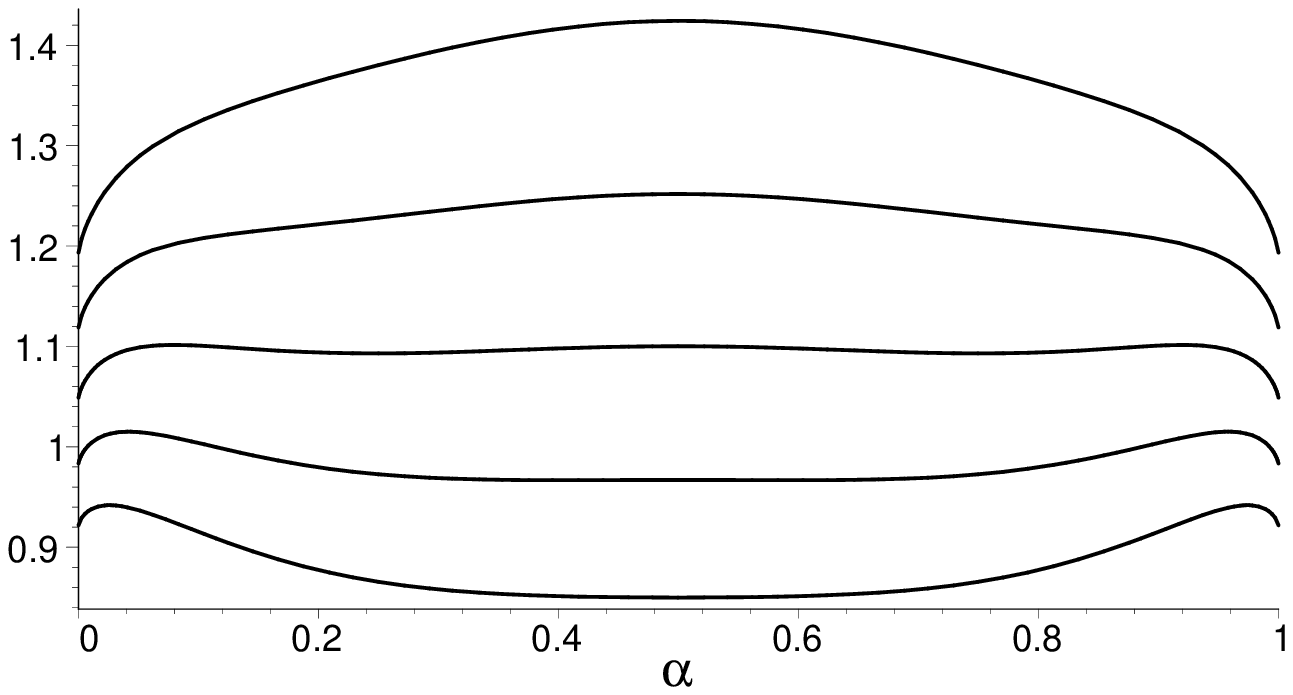,height=1.5in,width=2in}}
  \centerline{$k=5$, $r=8,9,10,11,12$ (top to bottom)}
\bigskip
  \centerline{\psfig{figure=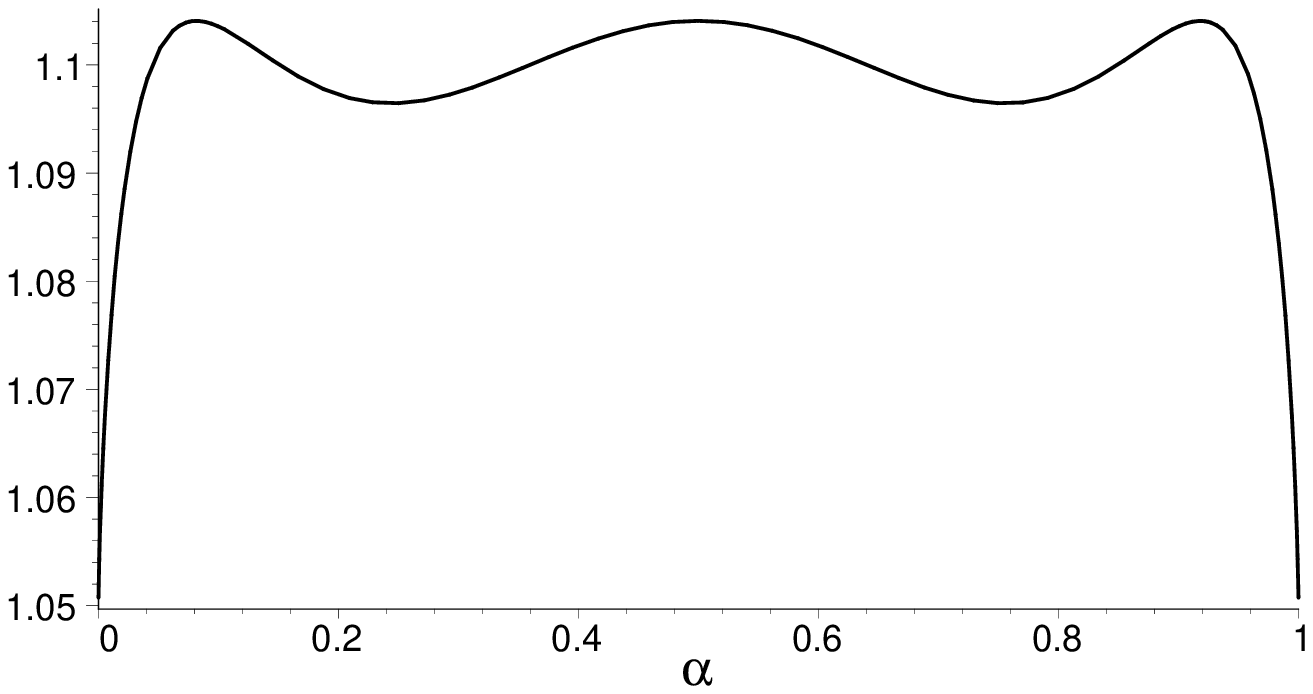,height=1.5in,width=2in}}
  \centerline{$k=5$, $r=9.973$}
\bigskip
Figure 2.  {\em The $n$th root of\/ $2 \times$the
entropy-correlation product for NAE $k$-SAT as a function of the
overlap $\alpha=z/n$.}
\end{figure}

\vspace*{-0.6cm}

In Fig.~2 we plot the $n$th root of the entropy-correlation
product for NAE $k$-SAT for various values of $r$.  Let us start
with the top picture, where $k=5$ and $r$ increases from 8 to 12
as we go from top to bottom. For $r=8,9$ we see that, indeed, the
global maximum occurs at $\alpha=1/2$. As result, for such $r$ we
have $\ex[X^2] = \Theta(\ex[X]^2)$, implying that the formula is
NAE-satisfiable with positive probability.

For the cases $r=11,12$, on the other hand, we see that at
$\alpha=1/2$ the function has dropped below 1 and therefore
$\ex[X]^2 = o(1)$, implying that \whp\ $F_k(n,rn)$ has no
NAE-satisfying truth assignment. It is worth noting that for
$r=11$ we have $\Pr[X>0]=o(1)$, even though $\ex[X^2]$ is
exponentially large, due to the maxima close to $0$ and $1$.

The most interesting case is $r=10$ where $\alpha=1/2$ is a local
maximum (and greater than 1) but the global maxima occur at
$0.08,0.92$ where the function equals $1.0145$... (vs.\
$1.0023$...\ at $\alpha=1/2$).  Because of this, we have
$\ex[X^2]/\ex[X]^2 > (1.0144/1.0024)^n$, implying that the second
moment method only gives an exponentially small lower bound on
$\Pr[X>0]$ in spite of the fact that the expected number of
NAE-satisfying truth assignments is exponential. Note, also, that
according to Table~\ref{tab:val} the best known upper bound for
$k=5$ is $10.505>10$.

Indeed, the largest value for which the second moment succeeds for
$k=5$ is $r = 9.973...$ This is depicted in the bottom picture
where the three peaks have the same height. For $r>9.973$ the
peaks near 0 and 1 surpass the one at $\alpha=1/2$ and the second
moment method fails.

\Section{Intuition}\label{sec:new}

\SubSection{Reducing the variance}

Given two truth assignments $s,t$ that have overlap $z$ let
\[
{\mathrm{boost}}(z) = \frac{\Pr[\mbox{$t$ is satisfying} \mid
\mbox{$s$ is satisfying}]}{\Pr[\mbox{$t$ is satisfying}]}
\enspace .
\]
It is not hard to see that
\[
\frac{\ex[X^2]}{\ex[X]^2} \,=\, 2^{-n} \sum_{z=0}^n
\binom{n}{z} \,{\mathrm{boost}}(z) \enspace .
\]
To examine one particular source contributing to
${\mathrm{boost}}(z)$ in the case of random $k$-SAT, it is helpful
to introduce the following quantity: given a truth assignment $s$
and a formula $F$ let $Q=Q(s,F)$ be the total number of literal
{\em occurrences\/} in $F$ that are satisfied by $s$. Thus,
$Q(s,F)$ is maximized when $s$ assigns each variable its
``majority'' value.

It is well-known that, with respect to properties that hold \whp,
$F_k(n,m=rn)$ is equivalent to a random formula generated as
follows: first, for each literal $\ell$, generate $R_{\ell}$
literal occurrences, where the $\{R_{\ell}\}$ are i.i.d.\ Poisson
random variables with mean $kr/2$; then, partition these literal
occurrences randomly into $m$ parts of size $k$.

In this model we can easily factor $\Pr[\mbox{$s$ is satisfying}]$
as \mbox{$\sum_q\Pr[Q = q] \times \Pr[\mbox{$s$ is satisfying}
\!\mid\! Q=q ]$}. Clearly, the probability of $Q$ deviating
significantly from its expected value $km/2$ is exponentially
small. At the same time, though, any such increase in $Q$  affords
$s$ tremendous advantage in terms of its likelihood to be
satisfying. Moreover, since \whp\ each variable appears in $O(\log
n)$ clauses, this advantage will be very much shared with the
truth assignments having large overlap with $s$, thus contributing
heavily to the $\mathrm{boost}$ function and, as a result, to
$\ex[X^2]$.

On the other hand, if we consider the probability that $s$ is
NAE-satisfying it is clear that $s$ would like $Q$ to be as close
as possible to $km/2$. In other words, now the typical case is the
most favorable case and the clustering around truth assignments
that satisfy many literal occurrences disappears.

Whether this is the main reason for which the second moment method
succeeds for random NAE $k$-SAT remains an interesting question.
Considering {\em regular\/} random $k$-SAT, where all literals are
required to appear an equal number of times, seems like an
interesting test of this hypothesis.

\SubSection{Geometry and connections to
  statistical physics} \label{sec:replica}

Statistical physicists have developed a number of methods for
investigating phase transitions which, while non-rigorous, are often
in spectacular agreement with numerical and experimental results.  One
of these methods is the replica trick. The term ``replica'' comes from
the fact that when $q$ is an integer one can compute $\ex[X^q]$ by
considering the interactions between $q$ elementary objects, or
``replicas'', counted by $X$.  In our case, we consider two truth
assignments when calculating the second moment.

At a high level, the replica trick amounts to computing $\ex[\ln
X]$ by calculating $\ex[X^q]$ for all {\em integer\/} $q$ and then
plugging in the resulting formula to the expression $\ex[\ln X] =
\lim_{q\to 0}{(\ex[X^q] - 1)}/q$. The fundamental leap of faith,
of course, lies in allowing the analytic continuation $q \to 0$
from integer values of $q$. Even to get this far, however, one has
to deal with the often daunting task of computing $\ex[X^q]$ for
all integer $q$.

When $X$ counts objects expressed as binary strings, such as
satisfying assignments, to calculate $\ex[X^q]$ one must in
general maximize a function of $2^q-1$ ``overlaps'', each overlap
counting the number of variables assigned a given $q$-vector of
0/1 values by the $q$ assignments/replicas.  (Note that in random
[NAE] $k$-SAT, since variables are negated randomly in each
clause, we can take one of the $q$ assignments to be the all 0s,
so we only have $2^{q-1}-1$ overlaps.)

By taking another leap of faith, one can dramatically reduce the
dimensionality of this maximization problem to $q$ by assuming
{\em replica symmetry\/}, \ie that the global maximum is symmetric
under permutations of the replicas.  For satisfiability problems
this  means that all overlap variables with the same number of 1s
in their respective $q$-vector take the same value. While this
assumption is often wrong, it can lead to good approximations.  In
particular, replica symmetry was assumed in the work of Monasson
and Zecchina~\cite{MZRK} predicting $r_k \simeq 2^k \ln 2$.

A standard indicator of the plausibility of replica symmetry in a
given system is the (usually experimentally measured) distribution of
overlaps between randomly chosen ground states, in our case satisfying
assignments. If replica symmetry holds, this distribution is tightly
peaked around its mean; if not, \ie if ``replica symmetry breaking''
takes place, this distribution typically gains multiple peaks or
becomes continuous in some open interval.

Intriguingly, the second-moment method is essentially a calculation of
the overlap distribution in the {\em annealed approximation}, \ie
after we average over random formulas (giving formulas with more
satisfying assignments a heavier influence in the overlap
distribution).  For random NAE $k$-SAT we saw that, almost all the way
to the threshold, the overlap distribution is sharply concentrated
around $n/2$, since when we take $n$th powers the contribution of all
other terms vanishes.

In other words, we have shown that in the annealed approximation, the
overlap distribution behaves {\em as if the NAE-satisfying assignments
  were scattered independently throughout the hypercube.}

\Section{Groundwork}\label{ground}

Let $X$ be the number of NAE-satisfying assignments of
$F_k(n,m=rn)$. We start by calculating $\ex[X]$. For any given
assignment $s$, the probability that a random clause is satisfied
by $s$ is the probability that its $k$ literals are neither all
true nor all false. We call this probability $p = 1-2^{1-k}$.

Since clauses are drawn independently with replacement and we have
$m=rn$ clauses, we see that
\begin{equation}
 \ex[X] = \left( 2 p^r \right)^n \enspace .\label{eq:naefirst}
\end{equation}

To calculate $\ex[X^2]$ we first observe that, by linearity of
expectation, it is equal to the expected number of ordered pairs
of truth assignments $s,t$ such that both $s$ and $t$ are
NAE-satisfying. We claim that the probability that a pair of truth
assignments $s,t$ are both NAE-satisfying depends only on the
number of variables to which they assign the same value (their
overlap). In particular, we claim that if $s$ and $t$ have overlap
$z=\alpha n$, where $0 \leq \alpha \leq 1$, then a random
$k$-clause $c$ is satisfied by both $s$ and $t$ with probability
\begin{eqnarray}
 f(\alpha)
 & = & 1 \,-\, 2 \cdot 2^{1-k} \,+\, 2^{1-k}(\alpha^k + (1-\alpha)^k)
 \label{eq:expl} \\
 & = & 1 \,-\, 2^{1-k} \left(2-\alpha^k - (1-\alpha)^k  \right)
 \nonumber
 \enspace .
\end{eqnarray}
To see this, first recall that the probability of a clause $c$ not
being satisfied by $s$ is $1-p=2^{1-k}$. Moreover, if $c$ is not
satisfied by $s$, then in order for $c$ to also not be satisfied by
$t$, it must be that either all the variables in $c$ have the same
value in $t$ and $s$, or they all have opposite values. Since $s$ and
$t$ have an overlap of $z=\alpha n$ variables and the variables in
each clause are distinct, the probability of this last event is
$\alpha^k + (1-\alpha)^k$. Thus,~\eqref{eq:expl} follows by
inclusion-exclusion.

Now, since the number of ordered pairs of assignments with overlap
$z$ is $ 2^n \,\binom{n}{z} $ and since the $m=rn$ clauses are
drawn independently and with replacement we see that
\begin{equation}
 \ex[X^2] \,=\, 2^n \sum_{z=0}^n  \binom{n}{z} \,f(z/n)^{rn}
 \enspace .
\label{eq:naesum}
\end{equation}

We will bound this sum by focusing on its largest terms. The proof
of the following lemma, based on standard asymptotic techniques,
appears in the Appendix.
\begin{lemma}
\label{lem:peak} Let $F$ be a real analytic positive function on
$[0,1]$ and define $g$ on $[0,1]$ as
\[
g(\alpha) = \frac{F(\alpha)}
                 {\alpha^\alpha \,(1-\alpha)^{1-\alpha}} \enspace ,
\]
where $0^0 \equiv 1$. If there exists $\amax \in (0,1)$ such that
$g(\amax) \equiv \gmax > g(\alpha)$ for all $\alpha \neq \amax$, and
$g''(\amax) < 0$, then there exist constants $B, C > 0$ such that for
all sufficiently large $n$
\[
B \times \gmax^n
\,\le\, \sum_{z=0}^n \binom{n}{z} \,F(z/n)^n
\,\le\, C \times \gmax^n
\enspace .
\]
\end{lemma}

With Lemma~\ref{lem:peak} in mind we define
$$
g_r(\alpha) =
\frac{f(\alpha)^r}{\alpha^{\alpha}(1-\alpha)^{1-\alpha}} \enspace.
$$
We will prove that
\begin{lemma}\label{gleas}
For every $\eps > 0$, there exists $k_0=k_0(\eps)$ such
that for all $k \geq k_0$, if
\[ r \leq 2^{k-1} \ln 2 - \frac{\ln 2}{2} - \frac{1}{2} - \eps \]
then $g_r(\alpha) < g_r(1/2)$ for all $\alpha \neq 1/2$, and
$g_r''(1/2) < 0$.
\end{lemma}
Therefore, for all $r,k,\eps$ as in Lemma~\ref{gleas},
$$
\ex[X^2] < C \times (2g_r(1/2))^n \enspace ,
$$
where $C=C(k)$ is  independent of $n$. At the same time, observe
that $\ex[X]^2 = (4p^{2r})^n = (2g_r(1/2))^n$. Therefore, for all
$r,k,\eps$ as in Lemma~\ref{gleas},
$$
  \frac{\ex[X^2]}{\ex[X]^2}  <  C \enspace ,
$$
which, by Lemma~\ref{lemma:sec}, implies
\[
\Pr[X > 0] > 1/C \enspace .
\]
Thus, along with Corollary~\ref{cor:boost}, Lemma~\ref{gleas}
suffices to establish Theorems~\ref{thm:ksat} and \ref{thm:naeksat}.

\Section{Proof of Lemma~\ref{gleas}}\label{sec:gleas}

We wish to show that $g''_r(1/2) < 0$ and that $g_r(\alpha) <
g_r(1/2)$ for all $\alpha \ne 1/2$.  Since $g_r$ is symmetric
around $1/2$, we can restrict to $\alpha \in (1/2,1]$. We will
divide this interval into two parts and handle them with two
separate lemmata. The first lemma deals with $\alpha \in
(1/2,0.9]$ and also establishes that $g''_r(1/2) < 0$.
\begin{lemma}
\label{lem:nearhalf} Let $\alpha \in (1/2,0.9]$. For all $k \ge
74$, if $r \leq 2^{k-1} \ln 2$ then $g_r(\alpha) < g_r(1/2)$ and
$g_r''(1/2) < 0$.
\end{lemma}
The second lemma deals with $\alpha \in (0.9,1]$.
\begin{lemma}
\label{lem:far} Let $\alpha \in (0.9,1]$.  For every $\eps > 0$
and all $k \geq k_0(\eps)$, if $r \leq 2^{k-1} \ln 2 - \frac{\ln
2}{2} - \frac{1}{2} - \eps$ then $g_r(\alpha) < g_r(1/2)$.
\end{lemma}

Combining Lemmata~\ref{lem:nearhalf} and \ref{lem:far} we see that
for every $\eps > 0$, there exists $k_0=k_0(\eps)$ such that for
all $k \geq k_0$ if
\[ r \leq 2^{k-1} \ln 2 - \frac{\ln 2}{2} - \frac{1}{2} - \eps \]
then $g_r(\alpha) < g_r(1/2)$ for all $\alpha \ne 1/2$ and
$g_r''(1/2)<0$, establishing Lemma~\ref{gleas}.  We prove Lemmata
\ref{lem:nearhalf} and~\ref{lem:far} below. The reader should keep
in mind that we have made no attempt to optimize the value of
$k_0$ in Lemma~\ref{lem:far}, opting instead for proof simplicity.

\medskip

\noindent {\bf Proof of Lemma~\ref{lem:nearhalf}.} We will first
prove that for $k \geq 74$, $g_r$ is strictly decreasing in
$\alpha = (1/2,0.9]$, thus establishing $g_r(\alpha)<g_r(1/2)$.
Since $g_r$ is positive, to do this it suffices to prove that
$(\ln g_r)' = g'_r/g_r < 0$ in this interval. In fact, since
$g'_r(\alpha) = (\ln g_r)' = 0$ at $\alpha = 1/2$, it will suffice
to prove that for  $\alpha \in [1/2,0.9]$ we have $(\ln g_r)'' <
0$. Now,
\begin{eqnarray}
 (\ln g_r(\alpha))''
& = & r \left( \frac{f''(\alpha)}{f(\alpha)}
         - \frac{f'(\alpha)^2}{f(\alpha)^2} \right)
- \frac{1}{\alpha (1-\alpha)} \nonumber\\
& \leq & r \,\frac{f''(\alpha)}{f(\alpha)} - \frac{1}{\alpha
(1-\alpha)} \enspace . \label{eq:sec_der_b}
\end{eqnarray}
To show that the r.h.s.\ of~\eqref{eq:sec_der_b} is negative we
first note that for $\alpha \geq 1/2$ and $k > 3$,
\[ f''(\alpha) = 2^{1-k} k(k-1) (\alpha^{k-2} + (1-\alpha)^{k-2})
   < 2^{2-k} \alpha^{k-2} k^2 \]
is monotonically increasing. Therefore,
\[ f''(\alpha) \leq f''(0.9) < 2^{2-k} \,0.9^{k-2} \,k^2 \enspace .
\]

Moreover, for all $\alpha$, $f(\alpha) \geq f(1/2)= (1-2^{-k})^2$.
Therefore, since $1/(\alpha (1-\alpha)) \geq 4$ and $r \leq
2^{k-1} \ln 2$, it suffices to observe that for all $k \ge 74$,
\[ (2^{k-1} \ln 2) \times \frac{2^{2-k} \,0.9^{k-2} \,k^2}{(1-2^{-74})^2}
 - 4 < 0 \enspace .
\]

Finally, recalling that $g'(1/2) = 0$ and using
\[
(\ln g_r)'' =    \frac{g''(\alpha)}{g(\alpha)}
         - \frac{g'(\alpha)^2}{g(\alpha)^2}
\]
we see that at $g''_r(1/2)<0$ since $(\ln g_r)''(1/2)<0$.
% \mbox{}\hfill$\Box$
\bigskip

\noindent {\bf Proof of Lemma~\ref{lem:far}.}  By the definition
of $g_r$ we see that $g_r(\alpha) < g_r(1/2)$ if and only if
\begin{equation}
\label{neq:est}
        \left( \frac{f(\alpha)}{f(1/2)} \right)^r
    < 2 \alpha^{\alpha} (1-\alpha)^{1-\alpha} \enspace .
\end{equation}
Letting $h(\alpha) = - \alpha \ln \alpha - (1-\alpha) \ln
(1-\alpha)$ denote the entropy function, we see
that~\eqref{neq:est} holds as long as
\[ \frac{r}{\ln 2 - h(\alpha)} < \frac{1}{\ln (1+w)} \]
where
\[ w = \frac{f(\alpha) - f(1/2)}{f(1/2)} \enspace .
\]
Observe now that for $k > 3$, $f$ is strictly increasing in
$(1/2,1]$, so $w > 0$.  Moreover, for any $x>0$
\[
\frac{1}{\ln(1+x)} \geq \frac{1}{x}+\frac{1}{2}-\frac{x}{12}
\enspace .
\]
Since $f(\alpha)-f(1/2) = 2^{1-k} (\alpha^k+(1-\alpha)^k-2^{1-k})
< 2^{1-k}$ and $f(1/2) = (1-2^{1-k})^2 > 1 - 2^{2-k}$, we thus see
that it suffices to have
\begin{equation}
\label{eq:need}
  \frac{r}{\ln 2 - h(\alpha)}
< \frac{2^{k-1} - 2}{\alpha^k + (1-\alpha)^k - 2^{1-k}} +
\frac{1}{2} - \frac{2^{1-k}}{12} \enspace .
\end{equation}

Now observe that for any $0 < \alpha < 1$ and $0 \le q <
\alpha^k$,
\[ \frac{1}{\alpha^k - q}  \ge  1 + k(1-\alpha) + q \enspace . \]
Since $\alpha > 1/2$ we can set $q = 2^{1-k} - (1-\alpha)^k$,
yielding
\[
\frac{1}{\alpha^k + (1-\alpha)^k - 2^{1-k}} \,\ge\, 1 +
k(1-\alpha) + 2^{1-k} - (1-\alpha)^k \enspace .
\]
Since $2^k (1-\alpha)^k < 5^{-k}$, we find that~\eqref{neq:est}
holds as long as $r \leq \phi(y) - 2^{-k}$ where
\[
 \phi(\alpha) \equiv \bigl(\ln 2 - h(\alpha) \bigr)
 \!\! \left( 2^{k-1} + (2^{k-1}-2) k (1-\alpha) - \frac{1}{2} \right)
  .
\]

We are thus left to minimize $\phi$ in $(0.9,1]$. Since $\phi$ is
analytic its minima can only occur at $0.9$ or $1$, or where
$\phi'=0$. The derivative of $\phi$ is
\begin{eqnarray}
\label{eq:phiprime} \phi'(\alpha) & = & (2^{k-1} - 2)
\times \Biggl[ -k \,(\ln 2 - h(\alpha)) \\
+ & & \!\!\!\!\!\!\!\!\!\!\!\!(\ln \alpha - \ln (1-\alpha)) \left(
1 + k(1-\alpha) + \frac{3}{2^k-4} \right) \Biggr] \enspace .
\nonumber
\end{eqnarray}
Note now that for all $k>1$
\[ \lim_{\alpha \to 1} \phi'(\alpha) = - \frac{2^{k}-1}{2} \ln (1-\alpha) \]
is positively infinite.  At the same time,
\[ \phi'(0.9) < -0.07 \times 2^k k + 1.1 \,(2^k-1) + 0.3 \,k \]
is negative for $k \ge 16$.    Therefore, $\phi$ is minimized in
the interior of $(0.9,1]$ for all $k \geq 16$. Setting $\phi'$ to
zero gives
\begin{equation}
\label{eq:boot} - \ln (1-\alpha)
 = \frac{k \,(\ln 2 - h(\alpha))}
        {1 + k (1-\alpha) + 3/(2^k-4)}
 - \ln \alpha \enspace .
\end{equation}

By ``bootstrapping'' we derive a tightening series of lower bounds
on the solution for the l.h.s.\ of~\eqref{eq:boot} for $\alpha \in
(0.9,1)$.  Note first that we have an easy upper bound,
\begin{equation}
\label{eq:phiupper} - \ln (1-\alpha) < k \ln 2 - \ln \alpha
\enspace .
\end{equation}
At the same time, if $k > 2$ then $3/(2^k-4) < 1$, implying
\begin{equation}
\label{eq:philower}
 - \ln (1-\alpha)
 > \frac{k \,(\ln 2 - h(\alpha))}
        {2 + k (1-\alpha)}
 - \ln \alpha
\enspace .
\end{equation}
If we write $k(1-\alpha)=B$ then~\eqref{eq:philower} becomes
\begin{equation}\label{eq:contra}
- \ln (1-\alpha)
 >  \frac{\ln 2 - h(\alpha)}{1-\alpha}
  \left( \frac{B}{B+2} \right)
  - \ln \alpha \enspace .
\end{equation}

By inspection, if $B \geq 3$ the r.h.s.\ of~\eqref{eq:contra} is
greater than the l.h.s.\ for all $\alpha > 0.9$, yielding a
contradiction. Therefore, $k(1-\alpha) < 3$ for all $k > 2$. Since
$\ln 2 - h(\alpha) > 0.36$ for $\alpha > 0.9$, we see that for
$k>2$, \eqref{eq:philower} implies
\begin{equation}\label{eq:rissoto}
- \ln (1-\alpha) > 0.07 \,k \enspace .
\end{equation}
Observe now that, by~\eqref{eq:rissoto}, $k (1-\alpha) < k \,
\e^{-0.07 k}$ and, hence, as $k$ increases the denominator
of~\eqref{eq:boot} approaches $1$.

To bootstrap, we note that since $\alpha > 1/2$ we have
\begin{eqnarray}
 h(\alpha) & \le & -2 (1-\alpha) \ln (1-\alpha) \label{eq:ent_b_l}\\
           & < & 2 \,\e^{-0.07 \,k} (k \ln 2 - \ln 0.9) \label{eq:randkl} \\
           & < & 2 \,k \,\e^{-0.07 \,k} \nonumber
\end{eqnarray}
where~\eqref{eq:randkl} relies
on~\eqref{eq:phiupper},\eqref{eq:rissoto}. Moreover, $\alpha >
1/2$ implies $-\ln \alpha \le 2(1-\alpha) < 2 \,\e^{-0.07 \,k}$.
Thus, by using~\eqref{eq:rissoto} and the fact $1/(1+x)>1-x$ for
all $x > 0$, \eqref{eq:boot} gives for $k \geq 3$,
\begin{eqnarray}
 -\ln (1-\alpha) & > & \frac{k \,(\ln 2 - h(\alpha))}
                            {1 + k (1-\alpha) + 3/(2^k-4)} \nonumber \\
  & > & \frac{k \,(\ln 2 - 2 \,k \,\e^{-0.07 \,k}) }
             {1 + 2 \,k \,\e^{-0.07 \,k}} \nonumber \\
  & > & k \,(\ln 2 - 2 \,k\,\e^{-0.07 \,k}) (1 - 2 \,k \,\e^{-0.07 \,k} )
\nonumber \\
  & > & k \ln 2 - 4 \,k^2\,\e^{-0.07 \,k} \enspace . \label{eq:fried}
\end{eqnarray}
For $k \ge 166$, $4 \,k^2 \,\e^{-0.07 \,k} < 1$. Thus, by
\eqref{eq:fried}, we have $1 - \alpha < 3 \times 2^{-k}$. This, in
turn, implies $-\ln \alpha \leq 2(1-\alpha) < 6 \times 2^{-k}$ and
so, by \eqref{eq:ent_b_l} and~\eqref{eq:phiupper}, we have for
$\alpha > 0.9$
\begin{equation}\label{eq:entr_last}
h(\alpha) < 6 \times 2^{-k} (k \ln 2 - \ln \alpha)
          < 5 \,k \,2^{-k} \enspace .
\end{equation}

Plugging~\eqref{eq:entr_last} into~\eqref{eq:boot} to bootstrap
again, we get that for $k \geq 3$
\begin{eqnarray*}
 -\ln (1-\alpha) & > & \frac{k \,(\ln 2 - 5 \,k \,2^{-k})}
                            {1 + 3 \,k\,2^{-k} + 3/(2^k-4)} \\
 & > &  \frac{k \,(\ln 2 - 5 \,k \,2^{-k})}
             {1 + 6 \,k \,2^{-k}}\\
 & > & k \,(\ln 2 - 5 \,k \,2^{-k}) (1 - 6 \,k \,2^{-k})\\
  & > & k \ln 2 - 11 \,k^2 \,2^{-k}  \enspace .
\end{eqnarray*}
Since $e^x < 1+2x$ for $x < 1$ and $11 \,k^2 \,2^{-k} < 1$ for $k
> 10$, we see that for such $k$
\[ 1-\alpha < 2^{-k} + 22 \,k^2 \,2^{-2k} \enspace .
\]

Plugging into~\eqref{eq:phiupper} the fact $-\ln \alpha < 6 \times
2^{-k}$ we get $-\ln (1-\alpha) < k \ln 2 + 6 \times 2^{-k}$.
Using that $\e^{-x} \geq 1-x$ for $x\geq 0$, we get the closely
matching upper bound,
\[ 1-\alpha > 2^{-k} - 6 \times 2^{-2k} \enspace . \]

Thus, we see that for $k \ge 166$, $\phi$ is minimized at an
$\alpha_{\min}$ which is within $\delta$ of $1-2^{-k}$, where
$\delta = 22 \,k^2 \,2^{-2k}$. Let $T$ be the interval
$[1-2^{-k}-\delta, 1-2^{-k}+\delta]$. Clearly the minimum of
$\phi$ is at least $\phi(1-2^{-k}) - \delta \times \max_{\alpha
\in T} |\phi'(\alpha)|$. It is easy to see
from~\eqref{eq:phiprime} that if $\alpha \in T$ then
$|\phi'(\alpha)| \le 2 \,k \,2^k$.

Now, a simple calculation using that $\ln (1-2^{-k}) > - 2^{-k} -
2^{-2k}$ for $k \ge 1$ gives
\begin{eqnarray*}
 \phi(1-2^{-k})
& = & \frac{1}{2}
      \bigl( (2^k - k) \ln 2 + (2^k - 1) \ln (1-2^{-k}) \bigr) \\
& & \times \,\bigl( 1 + (k-1) \,2^{-k} - k \,2^{2-2k} \bigr) \\
& > & 2^{k-1} \ln 2 - \frac{\ln 2}{2} - \frac{1}{2} - k^2 \,2^{-k}
\enspace .
\end{eqnarray*}
Therefore,
\[ \phi_{\min}
 \ge 2^{k-1} \ln 2 - \frac{\ln 2}{2} - \frac{1}{2} - 45 \,k^3 \,2^{-k}
\enspace .
\]
Finally, recall that~\eqref{neq:est} holds as long as $r <
\phi_{\min} - 2^{-k}$, \ie
\[ r
 < 2^{k-1} \ln 2 - \frac{\ln 2}{2} - \frac{1}{2} - 46 \,k^3 \,2^{-k}
\enspace .
\]
Clearly, we can take $k_0=O(\ln \eps^{-1})$ so that for all $k
\geq k_0$ the error term $46 \,k^3 \,2^{-k}$ is smaller than any
$\eps > 0$.  
%\mbox{}\hfill$\Box$

\Section{Conclusions}\label{sec:conc}

We have shown that the second moment method can be used to to
determine the random $k$-SAT threshold within a factor of 2. We
also showed that it gives extraordinarily tight bounds for random
NAE $k$-SAT, determining the threshold for that problem within a
small additive constant.

At this point, it seems vital to understand the following:
\begin{enumerate}
\item Why does the second moment method perform so well for NAE $k$-SAT?  The
symmetry of this problem explains why the method gives a non-trivial
bound, but not why it gives essentially the exact answer.
\item How can we close the factor of 2 gap for the random $k$-SAT
threshold?  Are there other large subsets of satisfying assignments
that are not strongly correlated?
\item
Does the geometry of the set of satisfying assignments have any
implications for algorithms? Perhaps more modestly(?), is there a
polynomial-time algorithm that succeeds with positive probability
for $r= \omega(k) \,2^k/k$, where $\omega(k) \rightarrow \infty$?
What about $\omega(k)=\Theta(k)$?
\end{enumerate}

\medskip
\paragraph{Acknowledgements} We would like to thank Henry Cohn for
bringing~\cite{debruijn} to our attention and Chris Umans for
helpful comments regarding the presentation. Special thanks to
Paul Beame for reading earlier versions and making many helpful
suggestions.

\bibliographystyle{latex8}
\bibliography{stoc,extra,theory}

\appendix

\Section{Proof of Lemma~\ref{lem:peak}}

The idea is that because of the binomial coefficient, the sum only
has $\Theta(\sqrt{n})$ ``significant" terms, each of which is of
size $\Theta(\gmax^n / \sqrt{n})$. The proof amounts to replacing
the sum by an integral and then using the Laplace method for
asymptotic integrals~\cite{debruijn}.

We prove the upper bound first.  Recall the following form of
Stirling's approximation, valid for all $n > 0$:
\begin{eqnarray*}
n! & > & \sqrt{2 \pi n} \,n^n \,\e^{-n} \,\left(1+\frac{1}{12 n}
\right)\\
n! & < & \sqrt{2 \pi n} \,n^n \,\e^{-n} \,\left(1+\frac{1}{6 n}
\right)
   \enspace .
\end{eqnarray*}
Thus, for any $0 < z \leq n/2$, letting $\alpha=z/n$ we have
\begin{eqnarray}
\binom{n}{\alpha n} & < & \frac{1}{\sqrt{2 \pi n}}
\,\frac{1}{\sqrt{\alpha (1-\alpha)}}
     \,\left( \frac{1}{\alpha^\alpha \,(1-\alpha)^{1-\alpha}}
     \right)^n \nonumber \;\;\;\;\;\\
&  & \times \,\frac{1 + 1/(6n)}{1 + 1/(12z)} \nonumber \\
& \le & \frac{1}{\sqrt{2 \pi n}} \,\frac{1}{\sqrt{\alpha
(1-\alpha)}}
     \,\left( \frac{1}{\alpha^\alpha \,(1-\alpha)^{1-\alpha}}
     \right)^n
\label{eq:gleouras}
\end{eqnarray}
and, similarly, for any $0 < z < n$ we have
\begin{eqnarray}
\binom{n}{\alpha n} \!\!& > &\! \frac{1}{\sqrt{2 \pi n}}
\,\frac{1}{\sqrt{\alpha (1-\alpha)}}
     \,\left( \frac{1}{\alpha^\alpha \,(1-\alpha)^{1-\alpha}}
     \right)^n \nonumber \;\;\;\\
&  & \times \,\frac{1}{\bigl(1 + 1/(6z)\bigr)
                     \bigl(1 + 1/(6(n-z)\bigr)} \nonumber \\
& > & \frac{36}{49} \, \frac{1}{\sqrt{2 \pi n}}
\,\frac{1}{\sqrt{\alpha (1-\alpha)}}
     \,\left( \frac{1}{\alpha^\alpha \,(1-\alpha)^{1-\alpha}}
     \right)^n  . \nonumber
\end{eqnarray}

To prove the upper bound, we use~\eqref{eq:gleouras} to write
\begin{eqnarray}
\sum_{z=0}^n \binom{n}{z} \,F(z/n)^n & \leq & \frac{1}{\sqrt{2 \pi
n}} \sum_{0 < z < n}
      \frac{g(z/n)^n}{\sqrt{(z/n) (1-z/n)}} \nonumber \\
& & + \,F(0)^n + F(1)^n
\enspace . \label{eq:klop}
\end{eqnarray}
Let $\epsilon = \min\{\amax,1-\amax\}/2 > 0$.  Let $g_* < \gmax$
be the maximum value of $g$ in $[0,\epsilon]\cup[1-\epsilon,1]$.
Since $g_* \geq g(0) = F(0)$ and $g_* \geq g(1) = F(1)$, using
\eqref{eq:klop} we get
\begin{eqnarray}
\label{eq:sum}
& & \sum_{z=0}^n \binom{n}{z} \,F(z/n)^n \\
& \leq &  \frac{1}{\sqrt{2 \pi n}}
  \, \times \sum_{z = \eps n}^{(1-\eps) n}
  \frac{g(z/n)^n}{\sqrt{(z/n)(1-z/n)}}
  + n^{3/2} g^n_*
\nonumber \\
& \leq &  \frac{1}{\sqrt{2 \pi n}} \,
\frac{1}{\sqrt{\epsilon(1-\epsilon)}} \times \sum_{z = \eps
n}^{(1-\eps) n} g(z/n)^n + n^{3/2}  g^n_*
\enspace . \nonumber
\end{eqnarray}

Next, we wish to replace the sum in \eqref{eq:sum} with an
integral. We first recall that for any integrable function $\phi$
that is monotone in $[a,b]$
\begin{eqnarray*}
& & \left| \sum_{j=0}^s \phi\!\left(a+j\,\frac{(b-a)}{s}\right) -
\frac{s}{b-a} \int_a^b \phi(x) \,\dx \right| \\
& \leq & \max\{F(a),F(b)\} \enspace .
\end{eqnarray*}
Therefore if $\phi$ has $M$ extrema in $[a,b]$, we can divide
$[a,b]$ into $M+1$ intervals on which $\phi$ is monotone, giving
\begin{eqnarray}
& & \left| \sum_{j=0}^s \phi\!\left(a+j\,\frac{(b-a)}{s}\right) -
\frac{s}{b-a} \int_a^b \phi(x) \,\dx \right| \nonumber \\
& \leq & (M+1) \times \max_{a \leq x \leq b} \phi(x) \enspace .
\label{eq:integ}
\end{eqnarray}
Observe now that $y^n$ is a strictly increasing function of $y$ in
$[0,\infty)$ implying that $g^n$ is extremized at exactly the same
$\alpha \in [0,1]$ as $g$. Since $g$ is independent of $n$ and
analytic on the closed interval $[\eps,1-\eps]$, it follows that
it has at most $M$ extrema in $[\eps,1-\eps]$ for some constant
$M$, and therefore so does $g^n$ for all $n>0$. Finally, since
$\gmax > g(\alpha)$ for all $\alpha\neq\amax$ we get that for all
sufficiently large $n$, $\gmax^n > n^{3/2}g_*^n$. Thus,
using~\eqref{eq:integ}, we can rewrite \eqref{eq:sum} as
\begin{eqnarray}
& & \sum_{z=0}^n \binom{n}{z} \,F(z/n)^n \nonumber \\
& < & \frac{1}{\sqrt{2 \pi n}} \,
\frac{1}{\sqrt{\epsilon(1-\epsilon)}} \nonumber \\
& \times & \left( n \int_{\epsilon}^{1-\epsilon} g(\alpha)^n \,
\dalpha + (M+2) \,\gmax^n \right) \enspace . \label{eq:deal}
\end{eqnarray}
To deal with the integral in~\eqref{eq:deal} we will use the
Laplace method for asymptotic integrals.  The following lemma can
be found in~\cite[\S 4.2]{debruijn}:
\begin{lemma}\label{debr}
Let $h$ be a real continuous function. Assume that there exist
$x_0$ and $b,c>0$ such that: i) $h(x) < h(x_0)$ if $x \neq x_0$,
ii) $h(x) \leq h(x_0) - b$ if $|x-x_0|\geq c$, and iii) $h''(x_0)
< 0$.  If $\int_{-\infty}^{\infty} \e^{h(x)} \,\dx$ converges,
then for any $\varepsilon > 0$ and all sufficiently large $t$,
\[ \int_{-\infty}^{\infty} \e^{th(x)} \,\dx
 \, < \, \sqrt{\frac{2 \pi}{(-h''(x_0)-3\varepsilon) \,t}} \, \e^{th(x_0)}
\]
and there is a similar lower bound for any $\varepsilon < 0$.
\end{lemma}
To apply this lemma, we set $t=n$, and take any continuous $h$
such that $h(x) = \ln g(x)$ for $x \in (0,1)$, and such that
$h(x)$ goes to $-\infty$ as $|x| \to \infty$ sufficiently fast so
that $\int_{-\infty}^{\infty} \e^{h(x)} \,\dx$ converges.  Observe
that since $\ln y$ is strictly monotone in $[0,\infty)$, $h$ is
extremized at the same $x$ as $g$. Clearly, condition ii) of
Lemma~\ref{debr} is also satisfied and since $[\ln g(x)]'' =
g''(x)/g(x) - (g'(x)/g(x))^2$, we see that $h''(\amax) =
g''(\amax)/g(\amax) < 0$. Therefore, for all sufficiently large
$n$
\begin{eqnarray*}
& & \sum_{z=0}^n \binom{n}{z} \,F(z/n)^n \\
& < &
 \frac{1}{\sqrt{2 \pi n}} \,
\frac{1}{\sqrt{\epsilon(1-\epsilon)}} \\
& \times & \left( n \, \sqrt{\frac{2 \pi \gmax}{-g''(\amax) \,n} }
\, \gmax^n
     \,+\, (M+2) \,\gmax^n \right) \\
& = & C \times \gmax^n  \enspace .
\end{eqnarray*}

In order to prove the lower bound, again we take $\eps =
\min\{\amax,1-\amax\}/2
> 0$, and discard all the terms of the sum for which $\alpha
\notin [\eps,1-\eps]$.  Since $1/\sqrt{\alpha
  (1-\alpha)} \ge 2$, we have
\begin{eqnarray*}
 \sum_{z=0}^n \binom{n}{z} \,F(z/n)^n & \ge &
   \sum_{z=\eps n}^{(1-\eps)n} \binom{n}{z} \,F(z/n)^n \\
& > & 2 \,\frac{36}{49} \,\frac{1}{\sqrt{2 \pi n}}
    \,\sum_{z = \eps n}^{(1-\eps) n} g(z/n)^n  \enspace .
\end{eqnarray*}
Replacing this sum by an integral as before and using the lower
bound of Lemma~\ref{debr} gives
\begin{eqnarray*}
& & \sum_{z=0}^n \binom{n}{z} \,F(z/n)^n  \\
& > & \frac{72}{49} \,\frac{1}{\sqrt{2 \pi n}} \\
& & \times \,\left( n \, \sqrt{\frac{2 \pi \gmax}{-g''(\amax) \,n} }
\, \gmax^n
     \,-\, (M+1) \,\gmax^n \right) \\
& = & B \times \gmax^n  \enspace .
\end{eqnarray*}
%\mbox{}\hfill$\Box$
%
%In fact, with a little more work we can prove
%\[
%B \,=\, C \,=\,
%\frac{1}{\sqrt{\amax (1-\amax)}} \sqrt{\frac{\gmax}{-g''(\amax)}}
%\enspace .
%\]

\end{document}